\RequirePackage{fix-cm}
\documentclass[smallextended]{svjour3}       % onecolumn (second format)
\smartqed  % flush right qed marks, e.g. at end of proof
\usepackage{graphicx}
\usepackage{amssymb}
\journalname{Computational Particle Mechanics}
\begin{document}

\title{Consistent SPH simulations of the anisotropic dispersion of a contaminant
plume}
%\subtitle{Do you have a subtitle?\\ If so, write it here}

%\titlerunning{Short form of title}        % if too long for running head

\author{Jaime Klapp \and
        Leonardo Di G. Sigalotti \and 
	Carlos E. Alvarado-Rodr\'{\i}guez \and
	Otto Rend\'on
}

\authorrunning{Klapp et al.} % if too long for running head

\institute{Jaime Klapp (Corresponding author) \at
            Departamento de F\'{\i}sica, Instituto Nacional de Investigaciones
              Nucleares (ININ) \at
              Carretera M\'exico-Toluca km. 36.5, La Marquesa, 52750 Ocoyoacac, \at
              Estado de M\'exico, Mexico\\
              \email{jaime.klapp@inin.gob.mx}
           \and
	   Leonardo Di G. Sigalotti (Corresponding author) \at
	      Departamento de Ciencias B\'asicas, \at Universidad Aut\'onoma
              Metropolitana - Azcapotzalco (UAM-A) \at
              Av. San Pablo 180, 02200 Ciudad de M\'exico, Mexico\\
              Tel.: +52 55 21761287\\
              \email{leonardo.sigalotti@gmail.com}       
	   \and
	   Carlos E. Alvarado-Rodr\'{\i}guez \at
              Direcci\'on de C\'atedras CONACYT \at
              Av. Insurgentes Sur 1582, Cr\'edito Constructor, Benito Ju\'arez,
              03940 Ciudad de M\'exico, Mexico\\
              \emph{Present address:} Departamento de Ingenier\'{\i}a Qu\'{\i}mica DCNyE, \at
              Universidad de Guanajuato, Noria Alta S/N, 36000 Guanajuato, Mexico\\
              \email{carlos.alvarado@conacyt.mx}\\
           \and
           Otto Rend\'on \at
	      Centro de F\'{\i}sica, Instituto Venezolano de Investigaciones
	      Cient\'{\i}ficas (IVIC) \at
              Apartado Postal 20632, Caracas 1020-A, Venezuela\\
	      and\\
	      Departamento de Ciencias B\'asicas, \at Universidad Aut\'onoma
              Metropolitana - Azcapotzalco (UAM-A) \at
              Av. San Pablo 180, 02200 Ciudad de M\'exico, Mexico \\
	      \email{ottorendon@gmail.com}\\
}

\date{Received: date / Accepted: date}
% The correct dates will be entered by the editor

\maketitle

\begin{abstract}
Solute transport through heterogeneous porous media is governed by fluid advection,
molecular diffusion and anisotropic dispersion. The dispersion is assumed to obey
Fick's law and the dispersion coefficient is defined as a second rank tensor. 
However, this problem has revealed to be a very difficult one because independently
of the numerical methods employed, the solutions are seen to exhibit artificial
oscillations and negative concentrations when the dispersivity becomes anisotropic.
Here we report consistent SPH simulations of the anisotropic dispersion of a
Gaussian contaminant plume in porous media using the open source code DualSPHysics.
Consistency of the SPH method is restored by increasing the spatial resolution
along with the number of neighbours within the compact support of the interpolating
kernel. The solution shows that as the number of neighbours is increased with
resolution, full convergence of the numerical solutions is guaranteed regardless of
the dispersivity. However, despite the restored consistency, negative concentrations,
albeit at a lower level, are still present. This suggests that a compromise between
the number of neighbours and the size of the smoothing length must be guaranteed
such that sufficient implicit numerical diffusion remains to damp out unphysical
oscillations.
% \PACS{PACS code1 \and PACS code2 \and more}
% \subclass{MSC code1 \and MSC code2 \and more}
\end{abstract}

\section{Introduction}
\label{intro}

The diffusion of a fluid in a porous medium is often directionally dependent owing to its 
heterogeneity. In general, the dissemination of solutes in such media occurs via three 
different mechanisms: a) fluid advection, which causes the solute to move with the 
streamwise flow velocity, b) molecular diffusion, which causes the spreading of the solute 
due to local concentration gradients, and c) mechanical dispersion, which results in 
disordered velocity fields due to the tortuosity of the particle trails in the heterogeneous
medium. In other words, mechanical dispersion is caused by the different paths that the 
solute is constrained to take due to the random arrangement and interconnectivity of the
channels in the porous medium. Such heterogeneity leads to anisotropic dispersion of the 
fluid as the variability of solute concentration gradients is greatly enhanced in one 
particular direction.

Mechanical dispersion is modelled using Fick's law and anisotropic dispersion occurs
when the solute transport in the direction of the streamwise flow (longitudinal dispersion)
is at least an order of magnitude greater than that in the other two directions
perpendicular to the main flow (transverse dispersion) \cite{Bear1988}. The numerical
simulation of solute spreading due to the concurrent intervention of the above mechanisms
often relies on the solution of the classical advection-dispersion equation (ADE) coupled
to the continuity and momentum equations. The dispersion coefficient entering the ADE is
a second-rank tensor designed to accommodate molecular diffusion and dispersion of the
solute due to small-scale heterogeneity \cite{Bear1988}. This formulation allows to
calculate the interplay between the increase of the concentration gradients due to the
solute deformation by dispersion and their smoothing due to diffusion. However, the
capture of these processes has represented a challenge for any numerical method because
of the appearance of spurious oscillations and negative concentrations in the solutions.
The very rapid changes in magnitude and direction of the flow velocity in heterogeneous
media are in part responsible for this drawback. In traditional meshing schemes, the
difficulty arises when the dispersion is highly anisotropic and the main direction of
dispersion deviates from the mesh orientation
\cite{Potier2005,Nordbotten2005,Mlacnik2006,Yuan2008,Lipnikov2009,Arbogast2012,Kim2014}.

Parallel to grid-based schemes, the method of Smoothed Particle Hydrodynamics (SPH)
has also been used to simulate anisotropic dispersion
\cite{Herrera2009,Herrera2010,Herrera2013,Avesani2015,Tran2016}. SPH is a meshfree
Lagrangian method based on interpolation theory. In the last 30 years the method has
become very popular because of its robustness and ease of implementation. In SPH,
the fluid is represented by particles and the physical properties carried by a given
particle are determined from those of all neighbouring particles lying within the
range controlled by an interpolation function, more commonly called the smoothing kernel.
The collective motion of all particles then describes the flow pattern
\cite{Monaghan1992,Liu2003}. Compared to traditional numerical methods, SPH has the
property of reducing numerical diffusion when solving the ADE in highly heterogeneous
media \cite{Boso2013}. A further important advantage of SPH over grid-based schemes
is that its solutions are independent of the effects of grid orientation, which is one
of the main problems faced by Eulerian schemes \cite{Herrera2010}. In view of these
advantages, SPH appears to be better suited to simulate solute spreading and mixing
than grid-based methods. However, all these advantages are not enough to suppress
artificial oscillations and negative concentrations in the solutions when the dispersion
is highly anisotropic \cite{Herrera2009}. In particular, when standard SPH is applied
to anisotropic dispersive transport, the occurrence of negative concentrations 
accompanied by large errors and slow convergence rates is a common result. Since for
isotropic dispersion, the results are free from spurious oscillations that cause
negative values of the concentration regardless of the degree of disorder of the
particles \cite{Herrera2009,Herrera2013,Avesani2015,Alvarado2019}, it has been inferred
that the problem arises when the off-diagonal terms of the tensor dispersion coefficient
are nonzero, which is the case of anisotropic dispersion.

In addition to the work of Herrera and collaborators
\cite{Herrera2009,Herrera2010,Herrera2013}, efforts to minimize and eventually remove 
unphysical oscillations and negative concentrations in SPH simulations of anisotropic
dispersion have been made by Avesani et al. \cite{Avesani2015}, Tran-Duc et al.
\cite{Tran2016}, and more recently by Alvarado-Rodr\'{\i}guez et al. \cite{Alvarado2019}.
In particular, Avesani et al. \cite{Avesani2015} proposed a modified version of standard
SPH based on a Moving-Least-Squares Weighted-Essentially-Non-Oscillatory (MWSPH)
reconstruction technique on moving points. They found that compared to standard SPH,
the numerical solution improves even at high anisotropies of the local dispersion
tensor with negative concentrations limited to about $10^{-7}C_{0}$, where $C_{0}$ is
the initial concentration. Working in the same line, Tran-Duc et al. \cite{Tran2016}
reported accurate simulations of anisotropic dispersion based on a different SPH scheme,
called anisotropic SPH approximation for anisotropic diffusion (ASPHAD). In ASPHAD,
the diffusion operator is first approximated by an integral in a coordinate system in
which it is isotropic and then by means of an inverse transformation of the integral
the anisotropic character of the diffusion operator is recovered. Although this scheme
conserves the main diffusing directions, it is rather sensitive to particle disorder
and reduces the degree of anisotropy due to the SPH smoothing. Results using a
consistent SPH approach were further reported by Alvarado-Rodr\'{\i}guez et al.
\cite{Alvarado2019}. In this case, consistency is restored by means of scaling relations
that define the number of neighbours ($n$) and the smoothing length ($h$) in terms of
the total number of particles ($N$) and comply with the asymptotic limits $n\to\infty$
and $h\to 0$ when $N\to\infty$ for complete SPH consistency
\cite{Rasio2000,Read2010,Zhu2015,Sigalotti2016,Sigalotti2019}. In particular, it was
demonstrated by Read et al. \cite{Read2010} that zeroth-order errors will persist in
SPH calculations when working with a low number of neighbours even if
$N\to\infty$. The explicit functional dependence of the SPH discretization errors on
the SPH parameters was derived by Sigalotti et al. \cite{Sigalotti2019}, who found
that these errors are $\propto 1/n$. Thus the higher the number of neighbours, the
lower the discretization errors. Working with a million particles and 31590 neighbours,
Alvarado-Rodr\'{\i}guez et al. \cite{Alvarado2019} found convergence rates and
magnitudes of the negative concentrations comparable to those reported by Avesani et al.
\cite{Avesani2015} with their MWSPH method. These calculations showed that while
first-order consistency was achieved at the maximum employed resolution, this was not
enough to ensure a positive concentration everywhere.

In this paper, we use a consistent version of the DualSPHysics code to simulate the
anisotropic dispersion of a Gaussian contaminant plume to explore the level of
resolution that is necessary to further reduce or even remove the unphysical oscillations
that give rise to negative concentrations. The paper is organized as follows. In Section 2
we introduce the governing equations and describe the SPH formulation along with the
approach implemented to restore consistency. Section 3 describes the simulation test
model. The results and the relevant conclusions are given in Sections 4 and 5, respectively.

\section{Governing equations and numerical implementation}
\label{sec:1}

\subsection{Transport equations}

The solute transport in a heterogeneous medium is described by the ADE, which in
Lagrangian form can be written in terms of the equations
\begin{eqnarray}
\frac{dC}{dt}&=&\nabla\cdot (\mathbb{D}\cdot\nabla C)-C\nabla\cdot {\bf v},\\
\frac{d{\bf x}}{dt}&=&{\bf v},
\end{eqnarray}
for fluid dispersion and advection, respectively. Here $C=C({\bf x},t)$ is the solute
concentration, ${\bf v}={\bf v}({\bf x},t)$ is the fluid velocity and $\mathbb{D}$ is
the dispersion tensor defined by
\begin{equation}
D_{ij}=\left(\alpha _{\rm T}|{\bf v}|+D_{\rm m}\right)\delta _{ij}+
\left(\alpha _{\rm L}-\alpha _{\rm T}\right)\frac{v_{i}v_{j}}{|{\bf v}|},
\end{equation}
where $D_{\rm m}$ is the molecular diffusion coefficient, $\delta _{ij}$ is the
Kronecker delta, $\alpha _{\rm L}$ is the longitudinal dispersivity (in the direction
of the local flow velocity), $\alpha _{\rm T}$ is the transverse dispersivity (in the
direction orthogonal to the local flow velocity), the indices $i,j$ refer either
to the $x$ or $y$ direction in a two-dimensional rectangular system and
$|{\bf v}|=(v_{x}^{2}+v_{y}^{2})^{1/2}$.

\subsection{SPH formulation}

Equation (1) is written in SPH form as \cite{Herrera2013,Alvarado2019}
\begin{eqnarray}
\left(\frac{dC}{dt}\right)_{a}&=&\frac{1}{2}\sum _{b=1}^{n}\frac{m_{b}}{\bar\rho _{ab}}
{{\cal D}_{ab}}\frac{\left(C_{a}-C_{b}\right)}{|{\bf x}_{ab}|^{2}}{\bf x}_{ab}
\cdot\nabla _{a}W_{ab}\nonumber\\
&+&C_{a}\sum _{b=1}^{n}\frac{m_{b}}{\bar\rho _{ab}}\sum _{i=1}^{m}\left[
\left(v_{x_{i},a}-v_{x_{i},b}\right)\frac{\partial W_{ab}}{\partial x_{i,a}}\right],
\end{eqnarray}
where
\begin{equation}
{\cal D}_{ab}=\sum _{i=1}^{m}\sum _{j=1}^{m}\frac{4D_{ij,a}D_{ij,b}}{D_{ij,a}+D_{ij,b}}
\left[\Gamma\frac{x_{i,ab}x_{j,ab}}{|{\bf x}_{ab}|^{2}}-\delta _{ij}\right],
\end{equation}
$m=2$ in two-space dimensions, $n$ is the number of neighbours within the support of
particle $a$, ${\bf x}_{ab}={\bf x}_{a}-{\bf x}_{b}$,
$|{\bf x}_{ab}|^{2}={\bf x}_{ab}\cdot {\bf x}_{ab}$, $W_{ab}=W(|{\bf x}_{ab}|,h)$ is the
kernel function, $\bar\rho _{ab}=(\rho _{a}+\rho _{b})/2$ and $\Gamma =4$. For transport
in an external uniform velocity field, the convective term on the right-hand side of
Eq. (4) vanishes and the equation takes the form of a pure diffusion-dispersion
equation. In this case, coupling to the velocity field is only ensured through the
dispersion coefficient.

\subsection{Consistency considerations}

A drawback of standard SPH is its lack of particle consistency, which affects the
accuracy and convergence of the method. If a polynomial of order $p$ is exactly
reproduced by a SPH approximation, then the approximation is said to have 
$C^{p}$-consistency, or $(p+1)$th-order accuracy. In this regard, the issue of
consistency is related to how well the discrete equations can reproduce the exact
differential equations. However, in SPH the loss of particle consistency is due to
the discrepancy between the kernel and the particle approximation. For example, it
is well-known that standard SPH has $C^{0}$ and $C^{1}$ kernel consistency, which are
lost when passing from the kernel to the particle approximation. The inconsistency
arises because of zeroth-order truncation errors introduced by the SPH discretization
\cite{Read2010}. An error bound for both the kernel and particle approximations as
a function of the SPH interpolation parameters, namely the total number of particles
$N$, the smoothing length $h$ and the total number of neighbours $n$, was derived
by Sigalotti et al. \cite{Sigalotti2019} to be
\begin{equation}
{\rm Error}\leq\left(a_{0}+a_{1}h+a_{2}h^{2}\right)\frac{1}{n}+a_{2}^{(K)}h^{2},
\end{equation}
where only terms up to second order have been retained. The coefficients $a_{0}$,
$a_{1}$, $a_{2}$ and $a_{2}^{(K)}$ depend on the dimension, the interpolation
kernel, the radius of the kernel support and the derivatives of the function or
variable that is being approximated \cite{Sigalotti2019}. In Eq. (6), the term
proportional to $1/n$ is the contribution to the error from the particle
discretization, while the term proportional to $a_{2}^{(K)}$ is the contribution
from the kernel approximation. It is evident from this expression that the
particle approximation contributes with terms of zeroth-, first- and second-order
in $h$, while the kernel approximation is only second-order accurate. For large
values of $n$ (i.e., for large numbers of neighbours) the error is dominated by
the term $a_{2}^{(K)}h^{2}$, implying consistency. For small values of $h$ and
$n\sim 60-100$, as is usually assumed in SPH calculations, the error is dominated
by the zeroth-order term $a_{0}/n$. This term contributes with an irreducible 
error even when $N\to\infty$ and $h\to 0$, as was predicted by the analysis of
Read et al. \cite{Read2010}. Therefore, for finite values of $h$, as is often
the case in practical applications of SPH, consistency is restored only for
large numbers of neighbours, while sufficiently accurate results can be obtained
working with small values of $h$. This complies with the joint limit $N\to\infty$,
$h\to 0$ and $n\to\infty$ with $n/N\to 0$ established by Rasio \cite{Rasio2000}
and Zhu et al. \cite{Zhu2015} for complete consistency.

For large values of $n$ ($n\gg 1$), Zhu et al. \cite{Zhu2015} parameterized the
SPH error as $\sim 1/n^{\nu}$, where $\nu =0.5$ for randomly distributed
particles and $\nu =1$ for low-discrepancy sequences of particles as is more
appropriate for most applications of SPH. By combining this error with the leading
one of the kernel approximation ($\propto h^{2}$), they derived the scaling
relations $n\sim N^{1-3/\beta}$ and $h\sim N^{-1/\beta}$ for $\beta\in [5,7]$,
which satisfy the limit $h\to 0$ and $n\to\infty$ as $N\to\infty$. A value of
$\beta\approx 6$ is more appropriate if the smoothing is performed on irregularly
distributed particles so that $h\sim N^{-1/6}$ and $n\sim N^{1/2}$. For the
simulations of this paper we choose $h=N^{-1/6}$, which produces a family of
curves for the dependence of $n$ on $N$. Of all possible curves we have chosen
the scalings $n\approx 2.81N^{0.675}$ and $h\approx 1.29n^{-0.247}$, which provides
a reasonably good compromise between the size of $h$ and the computational speed.

To support large numbers of neighbours and maintain numerical stability, a
Wendland C$^{4}$ function is used for the kernel, which is defined as
\cite{Wendland1995,Dehnen2012}
\begin{equation}
W(q,h)=B(1-q)^{6}\left(1+6q+\frac{35}{3}q^{2}\right),
\end{equation}
if $q\leq 1$ and 0 otherwise, where $q=|{\bf x}-{\bf x}^{\prime}|/h$ and
$B=9/(\pi h^{2})$ in two-space dimensions.

\subsection{Time integration}

A predictor-corrector leapfrog scheme is implemented for time integration of the
particle positions and concentrations. In the predictor step quantities are updated
at an intermediate time level according to the prescriptions
\begin{eqnarray}
{\bf x}_{a}^{l+1/2}&=&{\bf x}_{a}^{l-1/2}+\Delta t{\bf v}_{a}^{l},\nonumber\\
C_{a}^{l+1/2}&=&C_{a}^{l}+\frac{1}{2}\Delta t\left(\frac{dC}{dt}\right)_{a}^{l}.
\end{eqnarray}
With these updates, the time rate of change of the concentration is calculated at the
intermediate time level for use in the corrector step where quantities are advanced to
the new time $t^{l+1}$ according to
\begin{eqnarray}
{\bf x}_{a}^{l+1}&=&{\bf x}_{a}^{l}+\Delta t{\bf v}_{a}^{l+1/2},\nonumber\\
C_{a}^{l+1}&=&C_{a}^{l}+\Delta t\left(\frac{dC}{dt}\right)_{a}^{l+1/2}.
\end{eqnarray}
For the simulation cases of this paper, the flow velocity is given as an external
constant input parameter and therefore ${\bf v}_{a}^{l}={\bf v}_{a}^{l+1/2}$.

According to Herrera and Beckie \cite{Herrera2013}, the timestep for numerical
stability follows by demanding that
\begin{equation}
\Delta t=C\min _{a}\left(\frac{h^{2}}{D_{xx,a}+D_{yy,a}}\right),
\end{equation}
where $C=0.1$.

\section{Test model}

A benchmark test which is amply used for testing the performance of numerical schemes
for highly anisotropic dispersion consists of releasing in a two-dimensional unbounded
domain a Gaussian plume of contaminant of mass $M=10000$ kg, whose initial concentration
is given by
\begin{equation}
C({\bf x},t=0)=C_{0}\exp\left[\frac{-(x-x_{0})^{2}-(y-y_{0})^{2}}{2w^{2}}\right],
\end{equation}
where $C_{0}=0.32$ kg m$^{-3}$ is the maximum initial concentration at the centre of the
plume, $w=44$ m is the width of the Gaussian plume and ($x_{0}$,$y_{0}$) are the
coordinates of the plume centre at $t=0$ where the concentration equals $C_{0}$. The
flow velocity is assumed to be constant in space and time and the unbounded domain is
modelled by a square of side length $L=2000$ m. Periodic boundary conditions are applied
at the borders of the square and the contaminant is placed at the centre of the domain
($x_{0}=y_{0}=1000$ m). For a uniform velocity field, the solute concentration at any time
$t>0$ admits the analytical solution
\begin{equation}
\frac{C({\bf x},t)}{C_{0}}=\frac{w^{2}}{C_{4}}\exp\left[
        \frac{-(x-x_{0})^{2}A_{1}-(y-y_{0})^{2}A_{2}+4(x-x_{0})(y-y_{0})A_{3}}
                {8t^{2}C_{2}+4w^{2}tC_{3}+2w^{2}}\right],
\end{equation}
where
\begin{eqnarray}
        A_{1}&=&2tD_{yy}+w^{2},\nonumber\\
        A_{2}&=&2tD_{xx}+w^{2},\nonumber\\
        A_{3}&=&tD_{xy},\nonumber\\
        C_{2}&=&D_{xx}D_{yy}-D_{xy}^{2},\nonumber\\
        C_{3}&=&D_{xx}+D_{yy},\nonumber\\
        C_{4}&=&\left(4t^{2}C_{2}+2tw^{2}C_{3}+w^{4}\right)^{1/2}.
\end{eqnarray}
The molecular diffusion coefficient $D_{\rm m}$ is set to zero, the longitudinal
dispersivity is $\alpha _{\rm L}=10$ m and the constant flow velocity is
$v=1.16\times 10^{-5}$ m s$^{-1}$ so that the plume will travel about one metre per
day.
%---------------------------------------------------------Table 1
\begin{table}
\caption{Spatial resolution parameters.}
\label{tab:1}
\begin{tabular}{lllll}
\hline\noalign{\smallskip}
Number of SPH & Number of  & Mean particle & Smoothing & CPU time \\
particles     & neighbours & separation    & length    &   \\
\noalign{\smallskip}\hline\noalign{\smallskip}
$N$ & $n$ & $\Delta x/L$ & $h/L$ & $t(s)$ \\
\noalign{\smallskip}\hline\noalign{\smallskip}
1,000,000 & 31,529 & 0.00100 & 0.099 &  ~~98766.54 \\
2,000,000 & 50,339 & 0.00071 & 0.089 &  ~495064.78 \\
4,000,000 & 80,371 & 0.00051 & 0.079 &  1773900.15 \\
\noalign{\smallskip}\hline
\end{tabular}
\end{table}
%------------------------------------------------------------------

For the simulations we consider three different values of the dispersivity ratio,
namely $\alpha _{\rm T}/\alpha _{\rm L}=0.001$, 0.01 and 0.1. The flow velocity is
assumed to be oriented an angle $\theta =45^{\circ}$ measured with respect to the
$x$-axis and so $v_{x}=v_{y}=\sqrt{2}v/2$. The choice of these parameters allows direct
comparison with the standard SPH results of Herrera and Beckie \cite{Herrera2013},
the MWSPH simulations of Avesani et al. \cite{Avesani2015} and the lower resolution
calculations of Alvarado-Rodr\'{\i}guez et al. \cite{Alvarado2019}. The calculations
were performed with two and four million irregularly distributed particles and in all
cases they were terminated after 300 days. Initially, the number of neighbours and the
smoothing length are set using the scaling relations $n\approx 2.81N^{0.675}$ and
$h\approx 1.29n^{-0.247}$ in order to keep the discretization errors low and guarantee
$C^{0}$ and $C^{1}$ particle consistency (i.e., second-order accuracy). The spatial
resolution parameters for the simulations are listed in Table 1. According to the
above scaling relations, a run with 4 million particles will demand working with an
unprecedented number of neighbours ($n=80371$). For the three values
of $\alpha _{\rm T}/\alpha _{\rm L}$, the set of tests amounts to six independent runs.
The models were run using a modified version of the open-source code DualSPHysics
adapted for diffusive and dispersive transport problems \cite{Gomez2012,Crespo2015}.
The last column of Table 1 lists the CPU time in seconds employed by the simulations.
Good numerical accuracy is gained at the expense of an increased computational cost
owing to increasing the number of neighbours. As the resolution is doubled, the
number of neighbours increases by a factor of $\sim 1.6$ and the total CPU time to
by factors of $\sim 4-5$. The CPU times reported here is for non-uniformly distributed
particles within the domain of on an Intel(R) Xeon E5-2690 v3 CPU, with clockspeed
2.6 GHz and 12 cores.

\section{Results}

Although a dispersivity ratio $\alpha _{T}/\alpha _{L}=0.1$ is in line with most real 
situations \cite{Bear1988}, the performance of the numerical scheme is tested for even 
higher dispersivities ($\alpha _{T}/\alpha _{L}=0.01$ and 0.001). Similar anisotropy 
ratios were used by Herrera and Beckie \cite{Herrera2013}. A comparison of the
concentration fields with the analytical solution at different spatial resolutions is
displayed in Figs. 1, 2 and 3 after 300 days for $\alpha _{T}/\alpha _{L}=0.1$, 0.01 and 
0.001, respectively. The inset on the upper right side of each figure shows an amplified 
view of the numerical solution around the region of maximum concentration. For a flow 
orientation of $\theta =45^{\circ}$ the best match with the analytical profile is
observed for the lower dispersivity case ($\alpha _{T}/\alpha _{L}=0.1$) when working 
with $N=4,000,000$ particles and $n=80,371$ neighbours. Table 2 lists the 
root-mean-square errors (RMSEs) between the numerical and analytical concentration 
fields for varying resolution and anisotropy ratios.
%--------------------------------------------------------Table 2
\begin{table}
\caption{Root-mean-square errors between the analytical and numerical solution.}
\label{tab:1}
\begin{tabular}{lllll}
\hline\noalign{\smallskip}
Number of SPH & Number of  & RMSE & RMSE & RMSE \\
particles     & neighbours &      &      &      \\
\noalign{\smallskip}\hline\noalign{\smallskip}
$N$ & $n$ & $\alpha _{\rm T}/\alpha _{\rm L}=0.1$ & $\alpha _{\rm T}/\alpha _{\rm L}=0.01$ & 
$\alpha _{\rm T}/\alpha _{\rm L}=0.001$ \\
\noalign{\smallskip}\hline\noalign{\smallskip}
1,000,000 & 31,529 & $4.13\times 10^{-6}$ & $4.81\times 10^{-6}$ & $6.56\times 10^{-6}$ \\
2,000,000 & 50,339 & $2.38\times 10^{-6}$ & $4.17\times 10^{-6}$ & $5.53\times 10^{-6}$ \\
4,000,000 & 80,371 & $8.06\times 10^{-7}$ & $3.02\times 10^{-6}$ & $4.47\times 10^{-6}$ \\
\noalign{\smallskip}\hline
\end{tabular}
\end{table}
%-----------------------------------------------------------------

\begin{figure}
\includegraphics[width=0.8\textwidth]{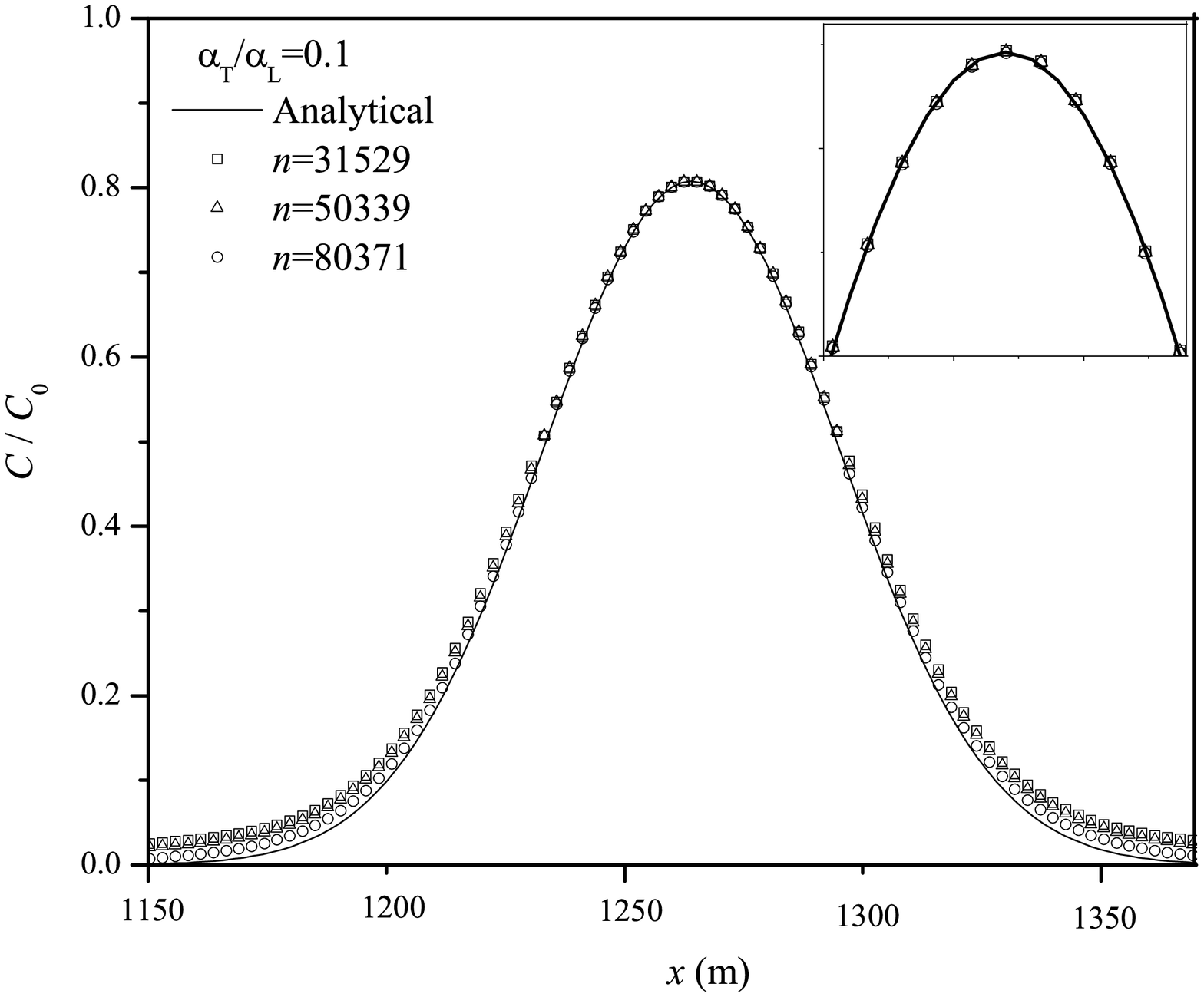}
\caption{Concentration profiles at different spatial resolutions normalized to
the initial concentration (symbols) as compared to the analytical solution given
by Eq. (12) (solid line) after 300 days for $\alpha _{\rm T}/\alpha _{\rm L}=0.1$.}
\label{fig:1}
\end{figure}

\begin{figure}
\includegraphics[width=0.8\textwidth]{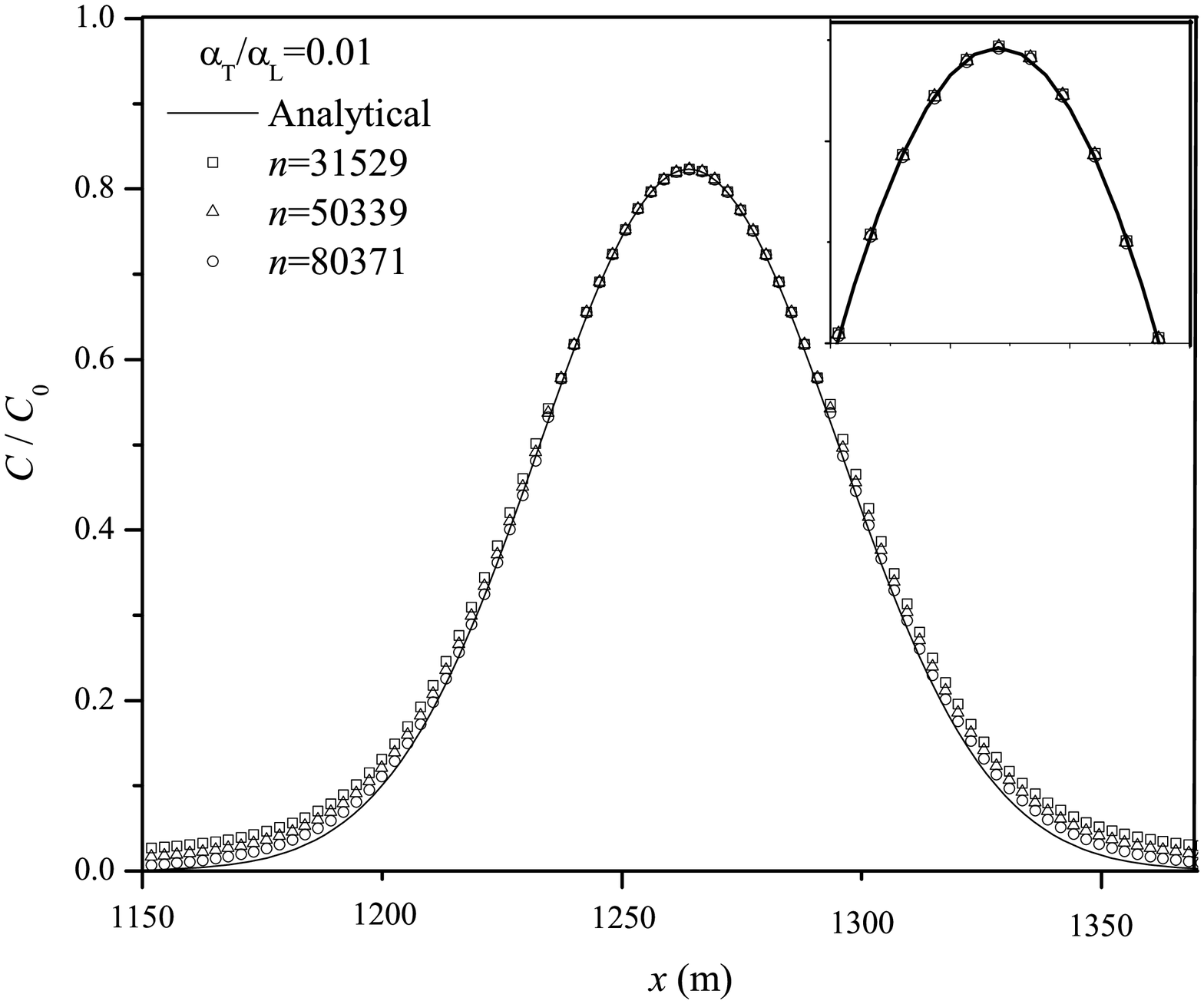}
\caption{Concentration profiles at different spatial resolutions normalized to
the initial concentration (symbols) as compared to the analytical solution given
by Eq. (12) (solid line) after 300 days for $\alpha _{\rm T}/\alpha _{\rm L}=0.01$.}
\label{fig:2}
\end{figure}

\begin{figure}
\includegraphics[width=0.8\textwidth]{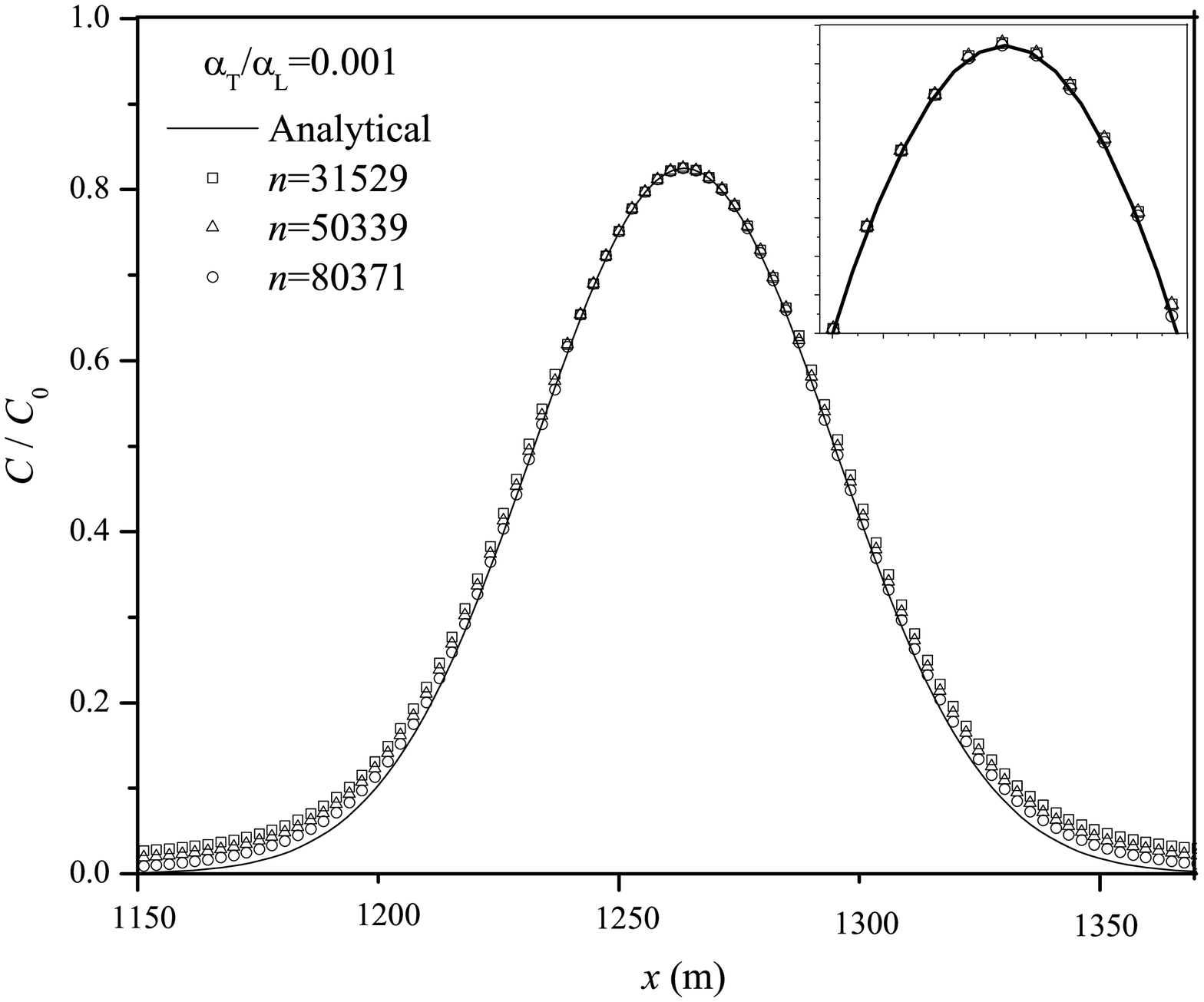}
\caption{Concentration profiles at different spatial resolutions normalized to
the initial concentration (symbols) as compared to the analytical solution given
by Eq. (12) (solid line) after 300 days for $\alpha _{\rm T}/\alpha _{\rm L}=0.001$.}
\label{fig:3}
\end{figure}

As the dispersivity increases from $\alpha _{\rm T}/\alpha _{\rm L}=0.1$ to 0.01 the
deviation of the numerical profile from the analytical one increases. However, at
the resolutions used here the differences in the RMSEs are small when doubling the
number of particles, suggesting that at relatively high spatial resolutions the
overall accuracy of the solution is almost independent of the dispersivity. As shown
in Figs. 1, 2 and 3 convergence to the analytical solution is already obtained for 
1,000,000 particles and 31,529 neighbours. When increasing the resolution to 2,000,000 
particles and 50,339 neighbours the deviation between the numerical and analytical 
profiles is reduced by about $1.8\times 10^{-4}$\% for 
$\alpha _{\rm T}/\alpha _{\rm L}=0.1$, and by $\sim 1.5\times 10^{-4}$\% when
passing from 2,000,000 particles and 31,529 neighbours to 4,000,000 particles and
80,371 neighbours. Similar percentages are obtained when increasing the resolution
for $\alpha _{\rm T}/\alpha _{\rm L}=0.001$. In all cases, the major contribution to
the error comes from the tail of the distribution when $C/C_{0}\to 0$, while it is
minimum around the peak of the distribution where the concentration is maximum. This
trends are consistent with the error bound in Eq. (6). If $h$ remains fixed, the
SPH discretization errors drop for sufficiently large numbers of neighbours and the
overall error is governed by the kernel approximation. This implies that consistency
is being restored. On the other hand, accuracy demands that $h$ must decrease as $n$ 
increases in which case the numerical solution approaches the analytical one. This
is in accordance with the scaling relations for which $h\sim n^{-0.247}$.
%---------------------------------------------------------------------Figure 4
\begin{figure}
\includegraphics[width=0.8\textwidth]{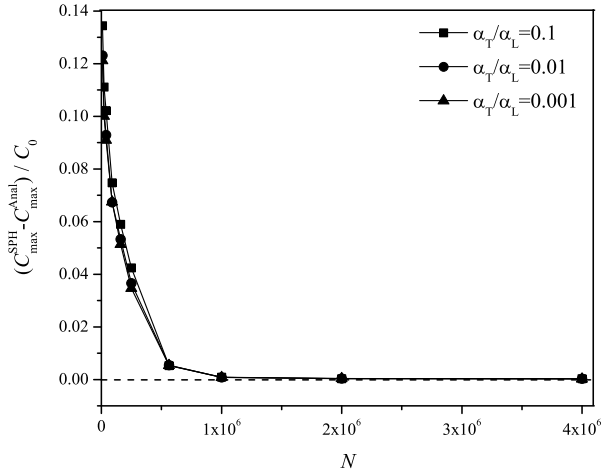}
\caption{Difference between the SPH and exact value of the maximum concentration
as a function of the total number of particles for the three dispersivity ratios
considered. Convergence is clearly achieved for $N>1,000,000$ particles. The 
data for $N<1,000,000$ particles correspond to the previous simulations by
Alvarado-Rodr\'{\i}guez et al. \cite{Alvarado2019}.}
\label{fig:4}
\end{figure}
%-----------------------------------------------------------------------------

A drawback of any numerical method when dealing with the problem of anisotropic
dispersion is the occurrence of negative concentrations away from the site of
the solute \cite{Herrera2009,Herrera2013,Avesani2015,Alvarado2019}. Part of the
problem lies on the tensor character of the dispersion coefficient given by
Eq. (3). That is, when the off-diagonal components of the dispersion coefficient
are nonzero, the solution is affected by artificial oscillations away from
the maximum concentration, which then amplify nonlinearly. As shown in Figs. 1,
2 and 3, any small oscillation about the zero in the tail of the distribution
(where $C/C_{0}\to 0$) can induce negative values of the concentration. This
problem has posed severe limitations for the correct prediction of anisotropic
solute dispersion and diffusion. As was pointed out by Compte and Metzler
\cite{Compte1997} the problem is related to the nature of the classical Fick's
law given by Eq. (1), which due to its parabolic character is endowed with an
infinite velocity of propagation. This can be seen from the analytical solution
(12), where for any time a finite amount of diffusing contaminat will always be 
present at very large distances from the solute (i.e., in the tail of the
concentration distribution), implying an infinitely fast propagation. While a
solution to this problem goes through converting the Fick's law into the hyperbolic
Cattaneo's equation \cite{Cattaneo1948}
\begin{equation}
\tau\frac{d^{2}C}{dt^{2}}+\frac{dC}{dt}=\nabla\cdot (\mathbb{D}\cdot\nabla C)-
C(\nabla\cdot {\bf v}),
\end{equation}
where the term $ \tau d^{2}C/dt^{2}$ is added \textit{ad hoc} to force a finite
propagation velocity along the transverse and longitudinal directions over a
characteristic time constant $\tau$, here we explore the ability of the present
scheme to suppress negative concentrations as the SPH consistency is restored by
increasing the number of neighbours. Previous consistent SPH simulations by
Alvarado-Rodr\'{\i}guez et al. \cite{Alvarado2019} have shown that convergence to
the analytical solution with the present scheme is already attained for 1,000,000 
particles and 31,529 neighbours, as shown in Fig. 4. It is evident from this figure 
that the difference between the SPH and the analytical solution tends to zero
independently of the ratio $\alpha _{\rm T}/\alpha _{\rm L}$, meaning that full
convergence is achieved when working with $N>1,000,000$ particles. However, this as 
not a sufficient condition to suppress the nonlinear amplification of small 
oscillations that are conducive to negative concentrations. In order to further explore 
under which conditions Eq. (1) can be used to model anomalous transport for large 
dispersivities with a consistent SPH scheme, we have increased the spatial resolution 
to 4,000,000 particles and 80,371 neighbours such that the ratio $n/N\approx 0.02$. In 
passing, we note that complete consistency will require $n/N\to 0$ for $N\to\infty$ and 
$n\to\infty$.

%---------------------------------------------------------------------Figure 5
\begin{figure}
\includegraphics[width=0.8\textwidth]{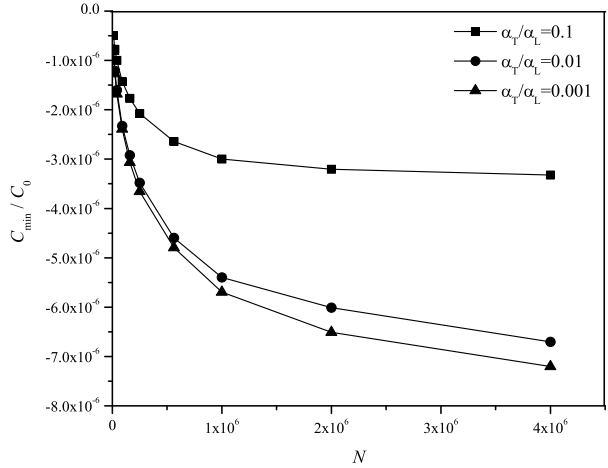}
\caption{Magnitude of the maximum negative concentration values normalized to
the initial concentration $C_{0}$ as a function of the total number of particles 
for all three dispersivity ratios. The data for $N<1,000,000$ particles correspond 
to the previous simulations by Alvarado-Rodr\'{\i}guez et al. \cite{Alvarado2019}.}
\label{fig:5}
\end{figure}
%-----------------------------------------------------------------------------

Figure 5 displays the maximum values of the negative concentration as a function
of the total number of particles for all three dispersivity ratios. As the
resolution is increased, the magnitude of the negative values increases as a
consequence of the decreased numerical diffusion which causes less damping of 
the unphysical oscillations. The smallest magnitudes occur for 
$\alpha _{\rm T}/\alpha _{\rm L}=0.1$. As the spatial resolution is increased, 
the rate of increase of the magnitude of the negative concentrations slows down.
For $N>1,000,000$ the maximum negative concentration tends asymptotically to a
constant value of $\approx -3.0\times 10^{-6}C_{0}$. At higher dispersivities 
the magnitude of the maximum negative concentration increases to 
$\approx -6.7\times 10^{-6}C_{0}$ (for $\alpha _{\rm T}/\alpha _{\rm L}=0.01$) and
$\approx -7.2\times 10^{-6}C_{0}$ (for $\alpha _{\rm T}/\alpha _{\rm L}=0.001$), 
respectively, with $N=4,000,000$ particles. However, there is no much difference 
in the trend and values of the negative concentrations when the dispersivity is 
increased from $\alpha _{\rm T}/\alpha _{\rm L}=0.01$ to 0.001. Compared to 
$\alpha _{\rm T}/\alpha _{\rm L}=0.1$, a constant value of the maximum negative 
concentration with resolution will require using a much higher number of particles
in these latter cases. Figures 6 depicts the concentration distributions at
different times during the spreading of the contaminant plume for 
$\alpha _{\rm T}/\alpha _{\rm L}=0.1$ (left mosaic) and 
$\alpha _{\rm T}/\alpha _{\rm L}=0.001$ (right mosaic). The white bands along the
transversal direction on both sides of the plume elongation represent the negative
concentrations. These bands always appear in the tail of the distributions where
the concentration decays asymptotically to zero.

%---------------------------------------------------------------------Figure 6
\begin{figure*}
\includegraphics[width=1.0\textwidth]{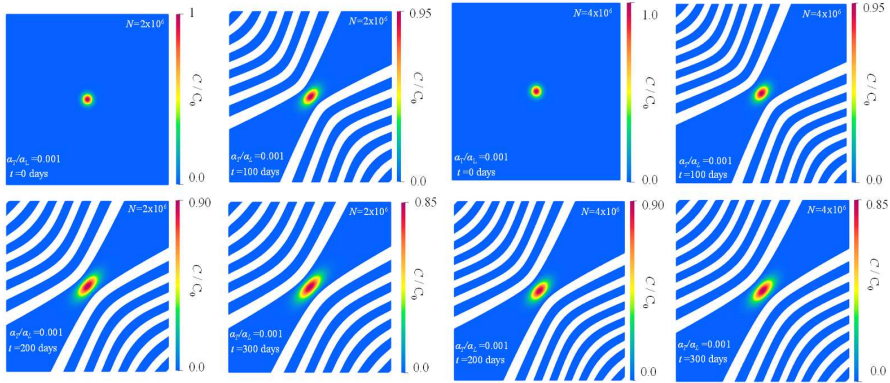}
\caption{Distribution of the concentration normalized to the initial value $C_{0}$
at different times during the plume spreading for
$\alpha _{\rm T}/\alpha _{\rm L}=0.001$. The left mosaic of frames corresponds to
the run with 2,000,000 particles and 50,339 neighbours, while that on the right to 
the run with 4,000,000 particles and 80,371 neighbours. Negative concentrations are 
evidenced by the white bands along the transversal direction on both sides of the 
elongated plume.}
\label{fig:6}
\end{figure*}
%-----------------------------------------------------------------------------

Standard SPH simulations by Herrera and Beckie \cite{Herrera2013} for the same test
case have produced maximum negative concentrations that are at least 4 orders of
magnitude higher than those reported here. The MWSPH scheme proposed by Avesani et
al. \cite{Avesani2015}, which is currently one of the best methods for the 
simulation of anisotropic dispersion and anomalous transport, produced absolute values 
of the negative concentrations less than about $10^{-7}C_{0}$ for 
$\alpha _{\rm T}/\alpha _{\rm L}=0.1$ and 0.01. In their case, the smoothing length of 
particle $a$ was defined according to the prescription 
$h_{a}=\sigma (m_{a}/\rho _{a})^{1/2}$ and the maximum absolute value of the negative 
concentrations was found to be sensitive to the value of $\sigma$ and the order $M$ of 
the Taylor series expansion of the concentration around the position $(x_{a},y_{a})$ of 
particle $a$, which was employed in the reconstruction procedure of the local concentration. 
For values of $M=3,4$ and $\sigma =3,2$, they reported magnitudes of the maximum negative 
concentrations of $10^{-12}-10^{-15}C_{0}$ for $\alpha _{\rm T}/\alpha _{\rm L}=0.1$ and 
$10^{-11}-10^{-13}C_{0}$ for $\alpha _{\rm T}/\alpha _{\rm L}=0.01$. On the other hand, 
according to Fig. 5 the almost constant behaviour of the $C_{\rm min}/C_{0}$-curve at 
high resolutions means that little can be improved by further increasing the resolution 
and the number of neighbours. That is, regardless of how much $N$ and $n$ are increased, 
the magnitude of the maximum negative concentrations will remain the same. According to 
Eq. (6), this suggests that for sufficiently large values of $n$ the global error will 
entirely depend on the kernel approximation through the last term in Eq. (6), which is 
$\propto h^{2}$. The sensitivity of the unphysical oscillations with the size of $h$ was 
further tested in an exploratory run with 4 million particles and using only 144 neighbours. 
This resulted in a smoothing length $h/L\approx 7.07\times 10^{-4}$ much smaller than those 
listed in Table 1. For such small value of $h$ the dominant term in Eq. (6) is the 
zeroth-order one given by $a_{0}/n$. Because of the excessive numerical diffusion
carried by this term, the simulation resulted in a very blurry concentration field and
was completely free of negative concentrations. The lesson to be learned from this result
is that the occurrence of unphysical oscillations can be minimized if the consistency
scaling relations are such that they provide smaller values of $h$ and correspondingly 
larger number of neighbours. This will introduce sufficient numerical diffusion
through the term $a_{0}/n$ to guarantee damping of the oscillations, while maintaining
reasonably good accuracy through small values of $h$. Unfortunately, the choice of the
best scaling relation must be done by trial and error.

%---------------------------------------------------------------------Figure 7
\begin{figure}
\includegraphics[width=0.8\textwidth]{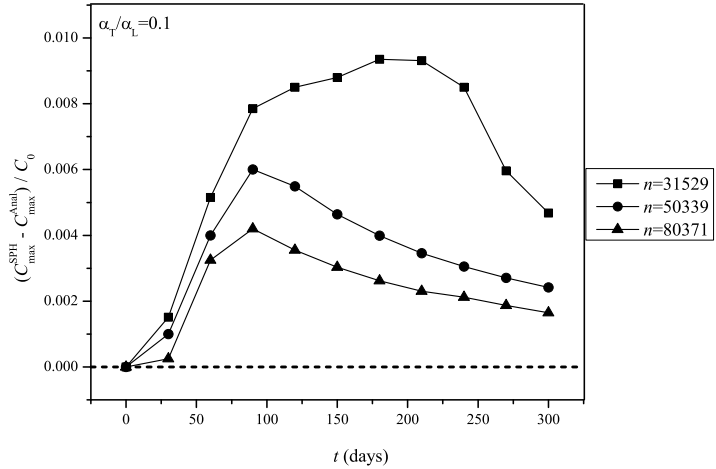}
\caption{Time evolution of the relative difference between the SPH and analytical     
maximum concentration values for $\alpha _{\rm T}/\alpha _{\rm L}=0.1$
and varying spatial resolution. The box on the right indicate the spatial
resolution for each curve in terms of the number of neighbours.}
\label{fig:7}
\end{figure}
%-----------------------------------------------------------------------------
%---------------------------------------------------------------------Figure 8
\begin{figure}
\includegraphics[width=0.8\textwidth]{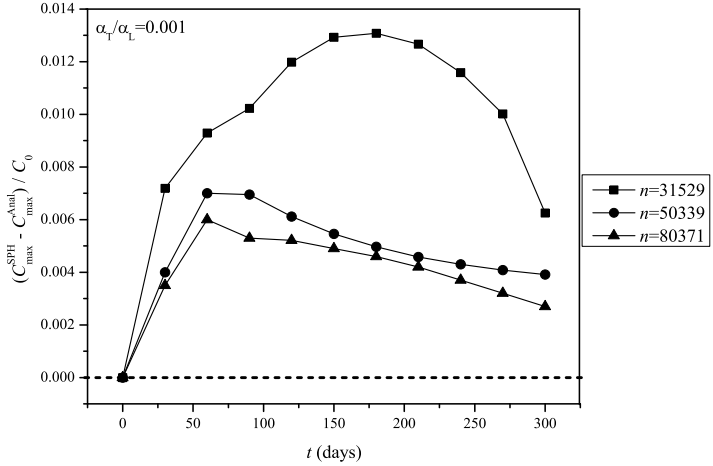}
\caption{Time evolution of the relative difference between the SPH and analytical 
maximum concentration values for $\alpha _{\rm T}/\alpha _{\rm L}=0.001$
and varying spatial resolution. The box on the right indicate the spatial 
resolution for each curve in terms of the number of neighbours.}
\label{fig:8}
\end{figure}
%-----------------------------------------------------------------------------

Figures 7 and 8 show the time variation of the relative difference between the 
analytical and numerical maximum concentrations for $\alpha _{\rm T}/\alpha _{\rm L}=0.1$
and 0.001, respectively. At early times, the relative differences are always less
than $\sim 0.01$ for $\alpha _{\rm T}/\alpha _{\rm L}=0.1$ and $\sim 0.013$
for $\alpha _{\rm T}/\alpha _{\rm L}=0.001$ with $N=1,000,000$ particles. These
differences drop to less than 0.4\% for $\alpha _{\rm T}/\alpha _{\rm L}=0.1$ and 
0.6\% for $\alpha _{\rm T}/\alpha _{\rm L}=0.001$ at the highest resolution. The 
positivity of the differences implies that the SPH simulations are overestimating 
the maximum concentration. At $t=300$ days, the realtive error decays to less than
0.3\% for the run with 4,000,000 particles regardless of the dispersivity. At early
times, these differences are comparable to those reported by Avesani et al. 
\cite{Avesani2015} with their MWSPH scheme for $\alpha _{\rm T}/\alpha _{\rm L}=0.1$ 
and irregularly distributed particles. In their case, however, negative differences 
were obtained at later times when working with $\sigma =3$ and $M=4$. The results
from both simulations agree that the shorter the smoothing length, the smaller the
error at all times. However, according to Eq. (6) such level of accuracy must be
accompanied by enough numerical diffusion to allow the unphysical oscillations to
damp out in the course of the contaminant spreading.

\section{Conclusions}

We have presented highly resolved, consistent SPH simulations of anisotropic dispersion 
of a solute in a heterogeneous porous medium using as a framework the DualSPHysics code. 
First-order consistency is restored by working with large numbers of neighbours within 
the kernel support. The number of neighbours, $n$, and the smoothing length, $h$, are set 
in terms of the total number of particles according to the power-law relations
$n\approx 2.81N^{0.675}$ and $h\approx 1.29n^{-0.247}$, which comply with the joint
limit $N\to\infty$, $n\to\infty$ and $h\to 0$ for complete consistency
\cite{Rasio2000,Zhu2015,Sigalotti2019}. The performance of the scheme is tested against the 
anisotropic spreading of a non-reactive, Gaussian solute plume that is instantaneously 
injected in a heterogeneous flow field for an irregularly distributed particles and
dispersivity ratios of $\alpha _{\rm T}/\alpha _{\rm L}=0.1$, 0.01 and 0.001
\cite{Herrera2013,Avesani2015,Alvarado2019}, where $\alpha _{\rm L}$ and $\alpha _{\rm T}$
are the longitudinal and transverse dispersivities, respectively. As the number of 
particles and neighbours is increased to unprecedented values, i.e., $N=4,000,000$ and 
$n=80,371$, the distance between the numerical and analytical solutions is highly reduced 
with root-mean-square errors that are less than $4.5\times 10^{-4}$\% in the worst case
of high dispersivity ($\alpha _{\rm T}/\alpha _{\rm L}=0.001$).

The numerical solutions exhibit convergence rates comparable to those reported by 
Avesani et al. \cite{Avesani2015} with their highly accurate MWSPH scheme. However, in
spite of the large number of neighbours used in the present simulations, the numerical
solutions are not free of unphysical oscillations in the tail of the solute 
concentration distribution, which are reponsible for induction of negative concentrations.
We find magnitudes of the maximum negative concentrations that are towards the upper 
limit of about $10^{-7}C_{0}$ reported by Avesani et al. \cite{Avesani2015}, where
$C_{0}$ is the maximum initial concentration. The results suggest that the above
scaling relations, which were chosen out of a family of infinite possible scalings
to guarantee a sharp drop of the particle discretization errors while keeping
reasonably large values of $h$ to alleviate the computational cost, was not enough to
suppress the nonlinear growth of numerical oscillations in regions away from the
contaminant plume, where the concentration tends asymptotically to zero. This is a
consequence of the numerically dispersive nature of the simulation which introduces 
little numerical diffusion. An exploratory run with 4,000,000 particles and a low
number of neighbours ($n=144$), which resulted in much smaller values of $h$ and much 
stronger diffusion due to the presence of irreducible zeroth-order discretization 
errors (because of the low number of neighbours), was free of negative concentrations 
away from the spreading plume, however, at the price of a very blurry concentration 
field everywhere due to the excessive numerical diffusion. Therefore, further consistent 
SPH simulations complying with the joint limit $N\to\infty$, $n\to\infty$ and $h\to 0$ 
must start working with scaling relations that guarantee a compromise between the number 
of neighbours and the size of $h$ such that sufficient numerical diffusion remains 
implicit in the scheme to damp out unphysical oscillations in regions where the 
concentration vanishes. 

\begin{acknowledgements}
The calculations of this paper were performed using the facilities of the ABACUS-Centro
de Matem\'atica Aplicada y C\'omputo de Alto Rendimiento of Cinvestav. We acknowledge
funding from the European Union's Horizon 2020 Programme under the ENERXICO Project,
grant agreement No. 828947 and from the Mexican CONACYT-SENER-Hidrocarburos
Programme under grant agreement B-S-69926. One of us (C.E.A.-R.) is a fellow 
commissioned to the University of Guanajuato under Project No. 368 and he acknowledges
finantial support from CONACYT under this project.
\end{acknowledgements}

\section*{Conflict of interest}

The authors declare that they have no conflict of interest.

\end{document}